\documentclass[a4paper]{jpconf}

\usepackage{verbatim}
\usepackage{graphicx}
\usepackage{dcolumn}
\usepackage{bm}
\usepackage{color}
\usepackage{url}
\usepackage{amsmath}
\usepackage{hyperref}

\newcommand{\be}{\begin{equation}}
\newcommand{\ee}{\end{equation}}
\newcommand{\ba}{\begin{eqnarray}}
\newcommand{\ea}{\end{eqnarray}}
\begin{document}

\title[title]{Astrophysics, cosmology, and fundamental physics with compact binary coalescence and the Einstein Telescope}

\author{C.~Van Den Broeck$^{1}$}
\address{$^1$Nikhef -- National Institute for Subatomic Physics, Science Park 105, 1098 XG Amsterdam, The Netherlands}
\date{\today}

\begin{abstract}
The second-generation interferometric gravitational wave detectors currently under construction are expected to make their first detections within this decade. This will firmly establish gravitational wave physics as an empirical science and will open up a new era in astrophysics, cosmology, and fundamental physics. Already with the first detections, we will be able to, among other things, establish the nature of short-hard gamma ray bursts, definitively confirm the existence of black holes, measure the Hubble constant in a completely independent way, and for the first time gain access to the genuinely strong-field dynamics of gravity. 
Hence it is timely to consider the longer-term future of this new field. The Einstein Telescope (ET) is a concrete conceptual proposal for a third-generation gravitational wave observatory, which will be $\sim 10$ times more sensitive in strain than the second-generation detectors. This will give access to sources at cosmological distances, with a correspondingly higher detection rate. I give an overview of the science case for ET, with a focus on what can be learned from signals emitted by coalescing compact binaries. Third-generation observatories will allow us to map the coalescence rate out to redshifts $z \sim 3$, determine the mass functions of neutron stars and black holes, and perform precision measurements of the neutron star equation of state. ET will enable us to study the large-scale structure and evolution of the Universe without recourse to a cosmic distance ladder. Finally, I discuss how it will allow for high-precision measurements of strong-field, dynamical gravity.  
\end{abstract}

%\maketitle

%--------------------------- INTRODUCTION --------------------------------------------------------%

\section{Introduction}
\label{sec:introduction} 

A network of second-generation, kilometer scale interferometric gravitational wave detectors is currently under construction: Advanced LIGO in the US \cite{LIGO}, Advanced Virgo in Europe \cite{Virgo}, and KAGRA in Japan \cite{KAGRA}; a further large interferometer might be built in India \cite{IndIGO}. In Germany, GEO-HF is operational \cite{GEO}. The first direct detection of gravitational waves (GWs) is expected before the end of the decade. Potential sources include pulsars, and indeed the (now decommissioned) first-generation detectors already put interesting bounds on the non-axisymmetry of the Crab pulsar \cite{Crab}. Next, there is the possibility of observing a stochastic gravitational wave background; here too, the initial observatories were already able to put relevant constraints \cite{Stochastic}. Among the most promising sources are coalescing compact binaries composed of neutron stars and/or black holes \cite{ratespaper}. Observing these will reveal the nature of short-hard gamma ray bursts (GRBs) \cite{GRBs}, establish once and for all the existence of black holes or uncover the presence of alternative compact objects such as boson stars \cite{Palenzuela07}, and enable us to independently study the expansion of the Universe by using coalescing binaries as `standard sirens' \cite{Schutz86}. They may also give us empirical access to the neutron star equation of state \cite{Hinderer09,Read09,Lackey11,Lackey13,Damour12,DelPozzo13,Yagi13}. Last but not least, they will allow us to probe the genuinely strong-field dynamics of spacetime  \cite{Yunes09,Mishra10,Cornish11,DelPozzo11,Li11a,Li11b,Chatziioannou12}; for a recent review, see \cite{VDB13}. Thus, an exciting field of observational gravitational physics is about to open up, with implications for astrophysics, cosmology, and fundamental physics.

It is appropriate to start considering the longer-term future of this new field. In 2008-2011, a conceptual design study was conducted for a third-generation gravitational wave observatory called Einstein Telescope (ET) \cite{ETdesign}. ET is envisaged to be a large ($\sim 10$ km), underground, partially cryogenic facility housing several interferometers arranged in a triangular configuration. It will be about a factor of ten more sensitive in strain than the second-generation detectors. While the second-generation observatories may see up to hundreds of sources per year \cite{ratespaper}, ET might have as many as a million detections. A first study of the considerable data analysis challenges associated with such a high detection rate was presented in \cite{Regimbau11}. In this paper, we will investigate in general terms what kind of novel science can be expected from ET. In doing so, we will not attempt to be exhaustive and focus on compact binary coalescence; for a more complete overview, see the recent Ref.~\cite{Sathya12}. 

This paper is structured as follows. In Sec.~\ref{sec:astrophysics}, we discuss the detailed reconstruction of the compact binary coalescence rate as a function of redshift which (a network of) third-generation detectors will allow for, the accuracy with which the mass function of  neutron stars and black holes can be determined, and the possibility of measuring the equation of state of neutron stars. In Sec.~\ref{sec:cosmology}, we turn to cosmology, in particular the potential of coalescing binaries as so-called standard sirens. Finally, in Sec.~\ref{sec:testingGR} we describe how high-precision tests of the strong field dynamics of general relativity could be performed with ET.  A summary is given in Sec.~\ref{sec:summary}.

Throughout this paper, we will use units such that $G=c=1$, unless stated otherwise. We will denote binary neutron stars by BNS, neutron star-black hole binaries by NSBH, and binary black holes by BBH.

\section{Astrophysics}
\label{sec:astrophysics}

\subsection{Reconstruction of the coalescence rate}

\begin{figure}[htbp!]
      \centering
	\includegraphics[angle=0,width=0.6\columnwidth]{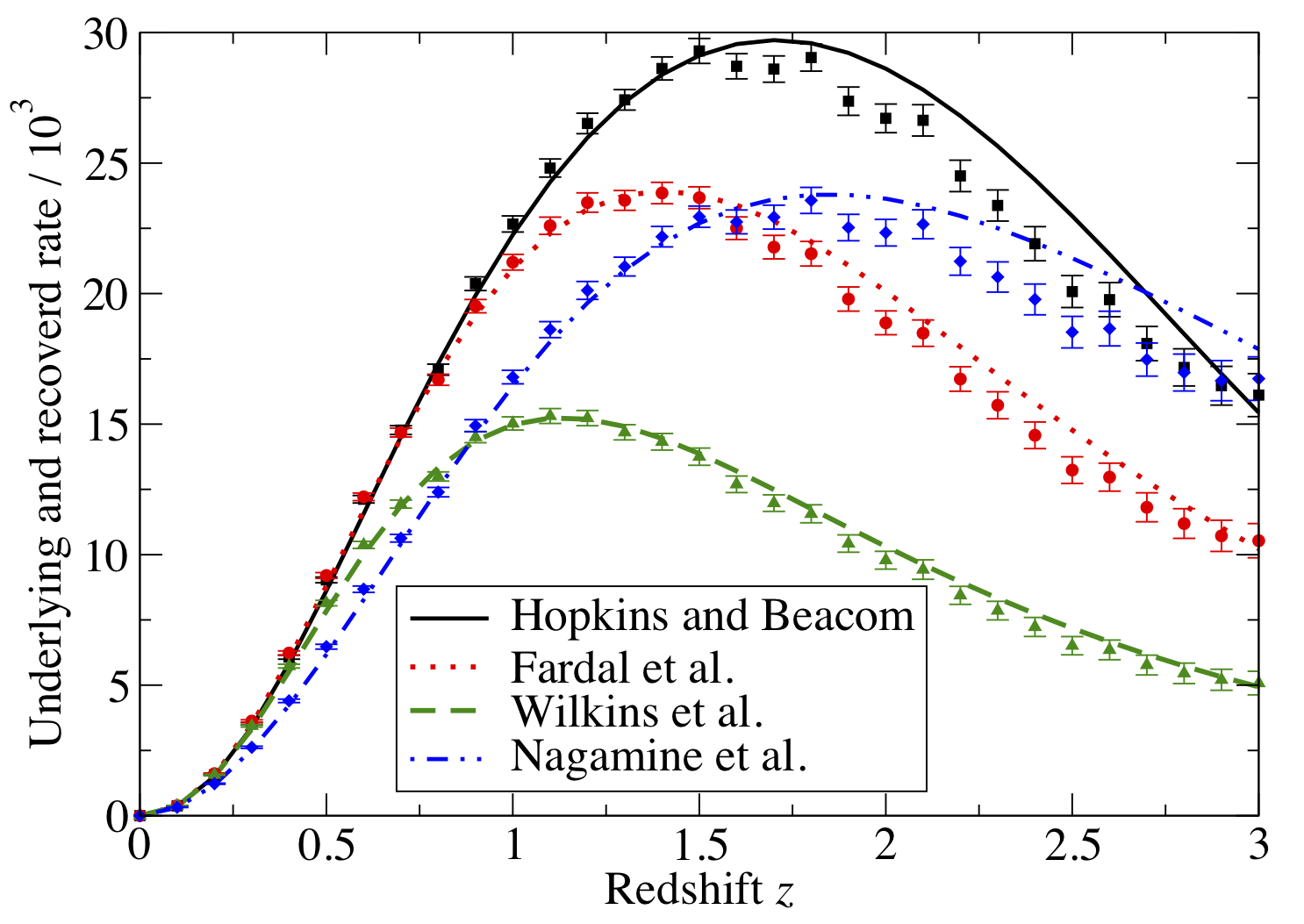}
	\caption{Four different models for the coalescence rate as a function of redshift, following Refs.~\cite{Hopkins06,Fardal07,Wilkins07,Nagamine06}. Assuming each of these in turn to be the underlying rate, one can simulate to what extent ET will be able to map it. The results are given by the `data points' with $1-\sigma$ error bars.}
	\label{fig:rates}
\end{figure}

ET will be able to see coalescing binaries out to cosmological distances. The coalescence rate as a function of redshift is expected to rise with increasing redshift $z$, reaching a peak around $z \sim 1.5$, after which it must decrease again and eventually reach zero at some large redshift value where no binaries had yet formed. The coalescence rate per redshift bin is given by \cite{Regimbau09}
\be
\frac{dR}{dz}(z) = \dot{\rho}_c(z) \frac{dV}{dz},
\ee
where $dV/dz$ is the derivative of comoving volume with respect to redshift, and
\be
\dot{\rho}_c(z) \propto \int_{t_d^{\rm min}}^\infty \frac{\dot{\rho}_\ast(z_f(z,t_d))}{1+z_f(z,t_d)}P(t_d)\,dt_d,
\ee
with $\dot{\rho}_\ast$ the star formation rate, $z$ the redshift at which merger takes place, $z_f$ the redshift at which the binary is formed,  and $P(t_d)$ a probability distribution of the delay time $t_d$ between formation and merger. The factor $(1+z_f)^{-1}$ within the integral takes into account the difference between intrinsic and observed rates. The advanced detectors, which will see mergers out to $z \simeq 0.3$, will be able to measure the coalescence rate density at $z=0$, $\dot{\rho}_c(0)$; even knowing this one number would already be valuable in understanding the fate of extremely massive stars. At larger redshifts, there is significant disagreement between various predictions for the \emph{shape} of the function $dR/dz(z)$, as illustrated in Fig.~\ref{fig:rates}, where coalescence rates were normalized such that $\dot{\rho}_c(0)$ is the same for all of them.

In order to arrive at a crude idea of how well a \emph{network} of ETs, \emph{e.g.} with locations at the two LIGO sites and the Virgo site, would be able to map the coalescence rate, one can proceed as follows. With such a network, one would be able to partially disentangle the luminosity distance $D_{\rm L}$ from the angles appearing in the GW amplitude (see Eq.~(\ref{amplitude}) below). Fisher matrix estimates indicate that distance could be estimated with $1-\sigma$ relative error $\sigma_0 \sim 1/\rho$, with $\rho$ the network SNR, if detector noise were the only source of uncertainty. However, distance measurements will also be affected by weak lensing. Modeling the associated relative distance error as $\sigma_w = 0.05\,z$ \cite{Sathya09}, one can take the combined uncertainty to be $\Delta D_{\rm L}/D_{\rm L} = (\sigma_0^2 + \sigma_w^2)^{1/2}$.  Next, consider a large number of simulated `catalogs' of $\mathcal{O}(10^5)$ sources each, distributed in redshift according to one of four different underlying true rate distributions $dR/dz(z)$ taken from recent literature \cite{Hopkins06,Fardal07,Wilkins07,Nagamine06}.  Within each catalog, take the `measured' distance for each source to be Gaussian distributed, centered around the true distance and with spread $\Delta D_{\rm L}/D_{\rm L}$. A measured distance $\hat{D}_{\rm L}$ can be converted into a measured redshift $\hat{z}$ using a cosmological model, \emph{e.g.} a $\Lambda$CDM cosmology with parameter values from the recent WMAP results \cite{WMAP}, and by counting the number of sources in each redshift bin one arrives at a `measured' coalescence rate as a function of redshift, for a given catalog. Starting from $z \sim 0.7$, this rate needs to be corrected for loss of GW detection efficiency due to inconvenient positioning and/or orientation of the source. Finally, in each redshift bin one can compute the average and variance of the measured $d\hat{R}/dz(z)$ over all the simulated catalogs. The results are shown as the `data points' and $1-\sigma$ error bars in Fig.~\ref{fig:rates}. For redshifts up to $z \sim 1.5$, the uncertainty on the rate is $\sim 1\%$. In particular, one would be able to tell the difference between the four theoretical distributions considered here.

\subsection{Reconstruction of the neutron star and black hole mass functions}

\begin{figure}[htbp!]
      \centering
	\includegraphics[angle=0,width=0.6\columnwidth]{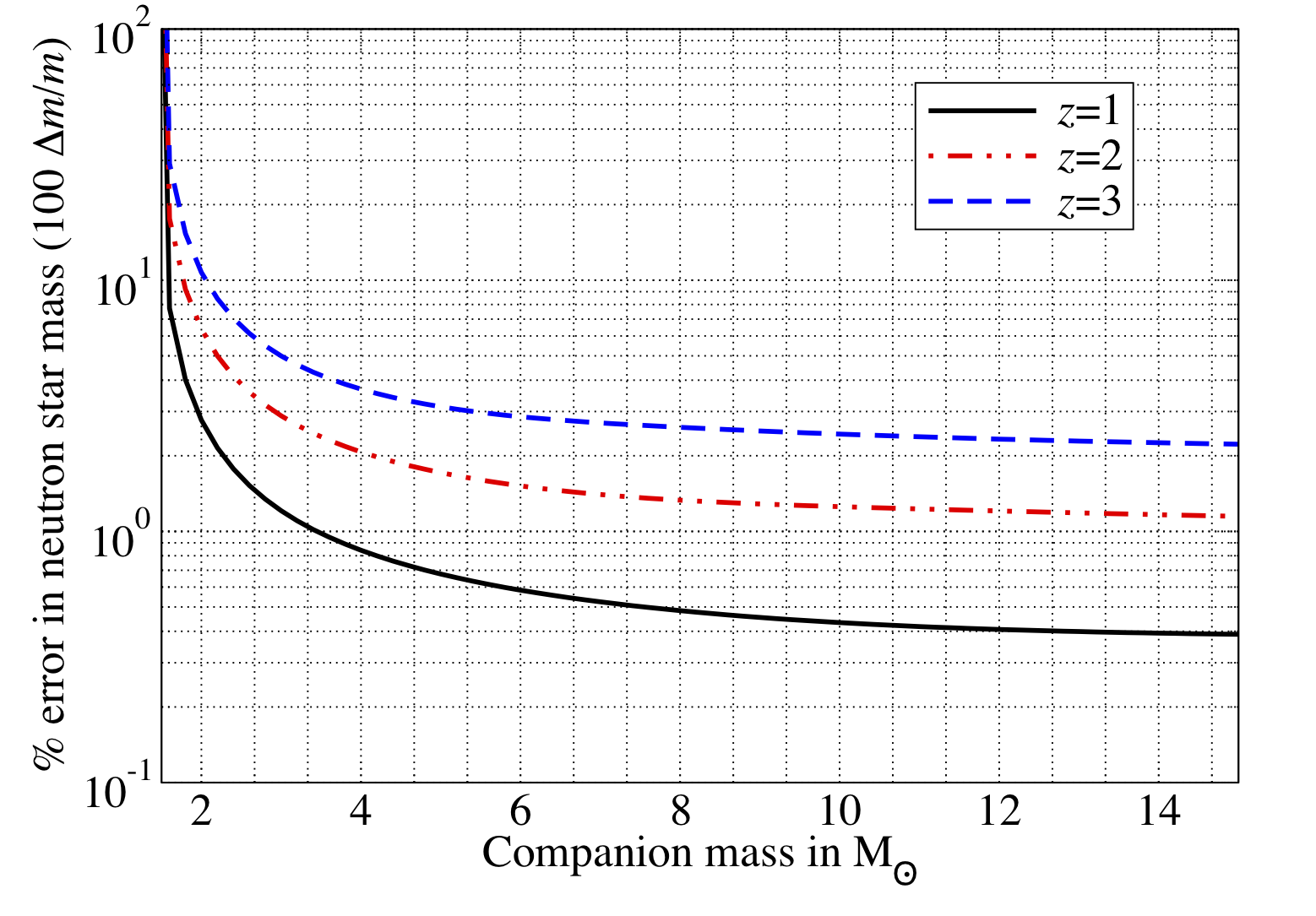}
	\caption{The percentage uncertainty in measuring the mass of a canonical ($1.4\,M_\odot$) neutron star in a BNS or NSBH system, as a function of the companion's mass, for sources at $z = 1, 2, 3$. Comparing with Fig.~\ref{fig:rates}, we see that ET would be able to measure component masses with $0.5-3\%$ uncertainty within the redshift interval where most of the inspirals take place.}
	\label{fig:NSmasses} 
\end{figure}

Current measurements of neutron star masses pertain to a small sample of neutron stars within our own galaxy (see \emph{e.g.} \cite{Ozel12}). Accordingly, the true distribution of neutron star masses, and in particular the mass region where the transition to black holes occurs, remains unknown. The possible time evolution of these mass functions over cosmological timescales is also something we currently have no access to. In Fig.~\ref{fig:NSmasses}, we show Fisher matrix estimates of how accurately the mass of a canonical ($1.4\,M_\odot$) neutron star in a BNS or NSBH system can be determined from the inspiral signal only, as a function of the companion's mass, and for $z = 1, 2, 3$. (The results suggests that in the near-equal mass case it will be difficult to estimate the mass; however, this is due to a singularity of the Fisher matrix itself.) As we have seen, star formation models suggest that the rate at which coalescences occur will peak at $z \simeq 1-3$, a range where ET can measure masses with high accuracy. Thus, ET will be able to make a detailed map of the neutron star and black hole mass functions during the most interesting part of star formation history. 

\subsection{Determining the equation of state of neutron stars}

The later part of BNS or NSBH inspiral, as well as the merger, will bear imprints of the neutron star equation of state. To linear order, the tidal field $\mathcal{E}_{ij}$ of the companion object will cause a spherically symmetric neutron star to acquire a quadrupole moment $Q_{ij} = -\lambda \mathcal{E}_{ij}$. The tidal deformability parameter $\lambda$ will be a function of the star's mass $m$. The neutron star's deformation affects the orbital motion and enters the inspiral waveform's phase starting from 5PN order (in the usual notation), where it contributes $\propto \lambda/M^5\,(v/c)^5$, with $M$ the total mass of the binary and $v$ the orbital velocity. Since $\lambda \propto R^5$, with $R$ the neutron star radius, and $R/M = \mathcal{O}(10)$ for a neutron star, the tidal contribution to the phase will be non-negligible despite occurring at such high order in $v/c$. The function $\lambda(m)$ is set  by the neutron star equation of state (EOS). 

In \cite{Hinderer09}, a variety  of EOS were considered, from very stiff to very soft, for different neutron star compositions (npe$\mu$ only, $\pi$/hyperon/quark matter, and strange quark matter). At $m = 1.4\,M_\odot$, the resulting values for $\lambda(m)$ vary over an order of magnitude.  Fisher matrix results indicate that in the gravitational wave frequency range 10 Hz - 450 Hz, with the advanced detector network and a single high-SNR source we will only have access to the stiffest EOS leading to the largest values of the deformability  $\lambda$. For a more recent assessment which also includes the plunge, see \cite{Damour12}. On the other hand, ET will have access to the entire range considered in \cite{Hinderer09}. 

In \cite{Read09}, merger and post-merger BNS waveforms were numerically computed for a range of EOS with varying stiffness. For very soft EOS, the merger is followed by the prompt formation of a black hole. By constrast, for a very stiff EOS, an unstable bar is formed, which spins down over a timescale of milliseconds before also collapsing to a black hole.  With second-generation detectors, in the post-merger it is again the stiffest equations of state that might cause a detectable signal because of the unstable bar. Thus, already with the upcoming advanced detector network, there is good hope of gaining information about the equation of state of neutron stars in this way. However, also in the post-merger phase, ET will enable far more accurate measurements due to better sensitivity in the kHz region, where the bar mode signal, if present, will have the most power. 

In \cite{Lackey11,Lackey13}, NSBH signals were considered, in which case larger induced neutron star quadrupole moments are to be expected. It was found that, for mass ratios of 2 or 3, advanced detectors would be able to extract $\lambda/M^5$ with 10-40\% accuracy for a single source at a distance of 100 Mpc, and in particular that they would be able to distinguish between a neutron star and a black hole of the same mass. ET would be able to do better by an order of magnitude.

Finally, as shown in \cite{DelPozzo13}, combining information from \emph{multiple} BNS inspirals observed with an advanced detector network will allow us to at least tell the difference between a soft, moderate, and stiff equation of state, and constrain the EOS to 10\% accuracy with as few as $\mathcal{O}(10)$ sources near the SNR threshold.

In summary, the prospects are good for empirically studying the EOS of neutron stars already with advanced detectors, but detailed measurements may have to wait until the ET era.

\section{Cosmology}
\label{sec:cosmology}

\subsection{Compact binary coalescences as standard sirens}

As noted by Schutz as early as 1986 \cite{Schutz86}, compact binary coalescences can be used as `standard sirens', a designation based on `standard candles', but referring to GW rather than electromagnetic sources. The observed strain amplitude of inspiral signals goes like
\be
\mathcal{A}(t) \propto \frac{1}{D_{\rm L}} \, \mathcal{M}_{\rm obs}^{5/3}\, \mathcal{F}(\theta,\phi,\iota,\psi)\,  F^{2/3}(t).
\label{amplitude}
\ee
Here $\mathcal{M}_{\rm obs}$ is the observed chirp mass (which is related to the \emph{physical} chirp mass $\mathcal{M}_{\rm phys}$ by $\mathcal{M}_{\rm obs} = (1+z)\,\mathcal{M}_{\rm phys}$), and $\mathcal{F}$ is a known function of sky position and orientation. The chirp mass can be measured very accurately from the phase. Thus, if some information is available on sky position and orientation, one can infer the luminosity distance $D_{\rm L}$ directly from the gravitational wave signal, with no need for calibration through a cosmic distance ladder, unlike the currently used standard candles. This opens up the possibility of completely independent cosmological measurements.

Assuming that at sufficiently large scales, the Universe can be modeled as a Friedman-Lema\^{i}tre-Robertson-Walker (FLRW) spacetime, its evolution is governed by the Hubble constant $H_0$, the density of matter $\Omega_{\rm M}$ relative to the critical density, the density of dark energy $\Omega_{\rm DE}$, the contribution from spatial curvature $\Omega_k$, and the dark energy equation of state parameter $w$. The latter is defined as $w = p_{\rm DE}/\rho_{\rm DE}$, with $\rho_{\rm DE}$ and $p_{\rm DE}$ the density and pressure of dark energy, respectively; note that we allow for the possibility that $w$ is time dependent. Given a number of sources for which the luminosity distance $D_{\rm L}$ can be measured, as well as the redshift $z$, the cosmological parameters $\vec{\Omega} = (H_0, \Omega_{\rm M}, \Omega_{\rm DE}, \Omega_k, w)$ can be inferred by fitting the distance-redshift relationship $D_{\rm L}(\vec{\Omega}; z)$.

\begin{figure}[htbp!]
      \centering
	\includegraphics[angle=0,width=0.6\columnwidth]{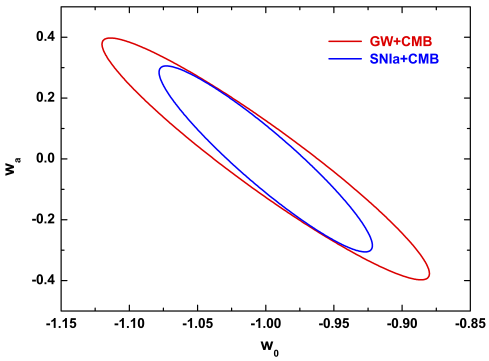}
	\caption{Measurement uncertainties for the possible time dependence in the dark energy equation of state parameter $w$, modeled as $w(a) \simeq w_0 + (1-a)\,w_a$, with $a$ the scale factor.  The slightly larger, red error ellipse is for standard sirens as seen with ET, the blue one for the possible future SNAP Type Ia supernova survey. In both cases, prior information from the CMB is assumed for $\Omega_{\rm M}$, $\Omega_{\rm DE}$, and $\Omega_k$. (Adapted from \cite{Zhao11}.)}
	\label{fig:w0wa} 
\end{figure}

This means that information about the redshifts of sources will also be required. In recent years, a variety of methods have been developed for utilizing binary inspiral signals as standard sirens. The first three will already allow for an independent measurement of $H_0$ with an advanced detector network.
\begin{enumerate}
\item Some fraction of inspirals will have electromagnetic counterparts, such as GRBs, which may allow for the identification of the host galaxy, and hence the redshift \cite{Nissanke09}.
\item Assuming knowledge of the distribution of \emph{physical} neutron star masses $m_{\rm phys}$, the \emph{observed} masses $m_{\rm obs} = (1+z)\,m_{\rm phys}$ in BNS sources lead to a redshift estimate without the need for electromagnetic counterparts \cite{Taylor11}.
\item Approximate knowledge of sky position by using multiple detector sites together with the distance uncertainty leads to a 3-dimensional volume within which the inspiral will have occurred. Using a galaxy catalog, each event can be associated with a list of possible redshifts, which for moderate distances can be translated into a list of possible values of $H_0$. A second inspiral event will lead to a different list of $H_0$ values which will only be partially consistent with the first. As more events are observed, the true value of $H_0$ will quickly emerge. This method allows the use of \emph{all} inspiral events, including BBH \cite{DelPozzo11b}.
\item If the equation of state of neutron stars has already been determined (\emph{e.g.} using some subset of the observed sources), one can extract the redshift directly from the GW waveform through the effect of tidal deformations on the orbital motion, which involves the \emph{intrinsic} masses \cite{Messenger11}. 
\end{enumerate}
In \cite{Sathya09} and \cite{Zhao11}, detailed studies were made of how accurately the cosmological parameters $\vec{\Omega}$ could be measured with the use of electromagnetic counterparts, \emph{i.e.} using the method (i) above. With $\mathcal{O}(1000)$ such events over the course of 5-10 years, $\Omega_{\rm M}$ and $\Omega_{\rm DE}$ could be measured with an accuracy comparable to what is possible with CMB measurements, but in a completely independent way. One can also consider the possible time dependence of the equation of state parameter $w$, taking the CMB measurements of  $\Omega_{\rm M}$, $\Omega_{\rm DE}$, and $\Omega_k$ as priors. It is convenient to approximate $w \simeq w_0 + (1-a)\,w_a$, with $w_0$ the present-day value, and $a$ the scale factor. In Fig.~\ref{fig:w0wa}, we compare accuracies in measuring $w_0$ and $w_a$, on the one hand using standard sirens seen by ET, and on the other hand considering the SNAP Type Ia supernova survey which may be available on the same timescale as ET \cite{SNAP}. The measurement quality is comparable in the two cases, but we stress once again that standard sirens allow for an \emph{independent} measurement; in particular, there will be no need for a cosmic distance ladder. More recent work by Taylor and Gair \cite{Taylor12} based on method (ii), which allows the use of \emph{all} BNS inspiral events seen in gravitational waves, has demonstrated that a network of ETs would in fact be able to beat the predicted accuracy of the future supernova surveys.  

\section{Probing the strong-field dynamics of spacetime}
\label{sec:testingGR}

Until the 1970s, all tests of general relativity (GR) took the form of observing test particles (photons, planets modeled as test masses, ...) moving in a weak, stationary gravitational field \cite{MTW}. This changed with the discovery of tight binary pulsars \cite{Hulse75,Burgay03,Kramer06}, whose orbital parameters were found to be changing in accordance with the emission of gravitational radiation as predicted by GR. Even so, the most interesting aspect of classical GR, namely the strong-field dynamics of the gravitational field itself, remains out of reach. In the foreseeable future, only the direct detection of gravitational radiation, in particular from coalescing compact binaries, will enable significant progress in this direction.

The authors of \cite{Arun06a,Arun06b,Mishra10} first pointed out the possibility of a very broad test of the dynamics of GR, based on a consistency requirement on the post-Newtonian coefficients appearing in the phase of an inspiral waveform. In the stationary phase approximation, the phase as a function of frequency takes the form \cite{Blanchet01,Blanchet04}
\be
\Psi(f) = 2\pi f t_c - \phi_c + \sum_{j=0}^7 \left[\psi_j + \psi_j^{(l)} \ln f \right]\,f^{(j-5)/3},
\label{phase}
\ee
where $t_c$ and $\phi_c$ are, respectively, the time and phase at coalescence. Now, according to general relativity, in the non-spinning case, all of the coefficients $\psi_j$ and $\psi_j^{(l)}$ only depend on the two component masses $(m_1,m_2)$, and GR makes a specific prediction for this dependence. Li \emph{et al.} \cite{Li11a,Li11b} subsequently developed a test of GR in the context of Bayesian inference, which computes an odds ratio based on the data $d$ and whatever background information $I$ one may hold:
\be
\mathcal{O}^{\rm modGR}_{\rm GR} = \frac{P(\mathcal{H}_{\rm modGR}|d,I)}{P(\mathcal{H}_{\rm GR}|d,I)}.
\label{oddsratio}
\ee 
Here $\mathcal{H}_{\rm GR}$ is the hypothesis that GR is correct, while $\mathcal{H}_{\rm modGR}$ is the hypothesis that \emph{one or more} of the $\psi_j$, $\psi_j^{(l)}$ do not have the functional dependence on component masses as welll as spins that GR predicts, without specifying which ones. Computing the posterior probability $P(\mathcal{H}_{\rm modGR}|d,I)$ amounts to testing a large number of independent sub-hypotheses, in each of which a particular subset of the phase coefficients are allowed to vary freely, while the others have the dependences on masses and spins as in GR. This has been dubbed the TIGER method. The framework is well suited to a situation where almost all sources have low SNR, as will be the case with the advanced detectors. It also trivially allows for combining information from multiple sources: for each source, the same `yes/no' question is asked, whose correct answer is the same for all sources, so that evidence for or against GR has a tendency to build up. 

As an example, one can consider a possible violation of GR in the coefficient $\psi_3$, which is of particular interest because 1.5PN is the lowest post-Newtonian order at which so-called `tail effects' become apparent, which roughly speaking result from the interaction of gravitational waves with the rest of spacetime \cite{Blanchet94,Blanchet95}.\footnote{This is an example of an essentially field-theoretic effect, rather than just pertaining to the behavior of test masses in a given gravitational field.  Such effects cannot be probed with the currently observed binary neutron stars, due to their low orbital velocity ($v/c \sim 10^{-3}$) and orbital compactness ($GM/c^2 R \sim 10^{-6}$) compared with a binary on the verge of merger (which has $v/c \sim 0.4$ and $GM/c^2 R \sim 0.2$, respectively). More generally, terms in Eq.~(\ref{phase}) with $j > 0$ are only accessible with direct gravitational wave detection.} As shown in \cite{Li11a}, with $\mathcal{O}(10)$ BNS detections, the Advanced LIGO and Virgo network will be able to pick up a 10\% deviation in this coefficient with essentially zero false alarm probability, and \emph{constraining} it will be possible at the $\sim 1\%$ level. In the case of ET, the Fisher matrix results of \cite{Mishra10} suggest that already with a \emph{single} source, we will be able to measure $\psi_3$ with an accuracy better than $0.1\%$ using the inspiral signal only. Presumably these numbers will be even better  for BBH when dynamical spins as well as merger and ringdown are taken into account.

It should be emphasized that a much wider class of GR violations can be picked up by using the TIGER method, not just simple shifts in post-Newtonian coefficients; see \cite{Li11a,Li11b} for some striking examples. A likely condition for a \emph{generic} violation to be detectable in this way is that, for fixed coalescence phase, it induces a shift in the waveform of half a cycle around the frequency where the detector is the most sensitive; both for the advanced detectors and for ET this is $\sim 150$ Hz. 
Finally, we note that although TIGER will be very efficient in \emph{finding} a deviation from GR, it will not necessarily allow for a determination of its precise nature. For the latter, one may want to use one of the general parameterized post-Einsteinian (ppE) waveform families of \cite{Yunes09,Cornish11,Chatziioannou12}.
%\footnote{We note that the general ppE waveforms considered in \cite{Cornish11} cannot be directly implemented in the TIGER framework. For instance, writing the phase as $\Psi(f) + \beta (\pi \mathcal{M} f)^b$, where $\Psi(f)$ is the GR prediction and $\beta$ and $b$ are free parameters, setting $\beta = 0$ but keeping $b$ free corresponds to the GR hypothesis, as does setting $b = 0$ but keeping $\beta$ free. This way, one loses the benefit of having many independent sub-hypotheses when dealing with large numbers of low-SNR sources \cite{Li11a}. Nevertheless, it will be of great interest to use such waveforms for parameter estimation on the highest-SNR events.} 

As shown in \cite{Gossan11}, observing the ringdown signal with ET will allow for a powerful test of the No Hair Theorem. As the black hole resulting from merger evolves towards a quiescent state, it is well modeled as a perturbed Kerr black hole. The Einstein equations dictate that the frequencies and damping times of the quasi-normal modes only depend on the mass $M$ and spin $J$ of the underlying Kerr geometry. For a $500\,M_\odot$ black hole at 6 Gpc, a departure as small as 8\% in the frequency of the dominant mode could be detected. Kamaretsos \emph{et al.} went a step further by relating ringdown mode amplitudes to the progenitor masses and spins \cite{Kamaretsos11,Kamaretsos12}. It would be of great interest to cast these ideas into the language of TIGER, in which case smaller deviations can presumably be found even for a single source, and information from multiple sources can be combined.

\section{Summary}
\label{sec:summary}

The advanced interferometric gravitational wave detectors that are currently being built have the potential to settle important scientific questions: What is the local coalescence rate, and what does this imply for our understanding of the evolution of massive stars? What is the nature of short-hard gamma ray bursts? Do black holes really exist, or is there some alternative kind of very compact object? What does the equation of state of neutron stars approximately look like? Are estimates of the present-day expansion of the Universe correct? Is general relativity really the correct theory of gravity in the strong-field, dynamical regime? 

Einstein Telescope will take us to much larger detection rates and out to cosmological distances, where new questions can be addressed. What is the mass distribution of neutron stars and black holes, and how did it evolve in time? What is the true equation of state of neutron stars (and did this, too, have time dependence)? In cosmology, we will arrive at a comprehensive, completely independent determination of all the basic parameters governing the Universe. Is the equation of state of dark energy time dependent? Last but not least, ET will probe the strong-field dynamics of spacetime in an unprecedented way.

\section*{Acknowledgements}

CVDB is supported by the research programme of the Foundation for Fundamental Research on Matter (FOM), which is partially supported by the Netherlands Organisation for Scientific Research (NWO). It is a pleasure to thank the organizers of ICGC 2011 for their kind invitation. The author would also like to thank the contributors to the Einstein Telescope Conceptual Design Study, and in particular B.S.~Sathyaprakash, for many fruitful discussions. 

\section*{References}
%\bibliographystyle{iopart-num.bst}
%\bibliographystyle{h-physrev.bst}
%\bibliographystyle{utphys.bst}
%\bibliographystyle{apsrev.bst}
%\bibliography{amaldi_proc_v1}

\end{document}